\documentclass[11pt,a4paper]{article}
\pdfoutput=1 

\usepackage{jheppub}

\usepackage{hyperref}
\usepackage{cleveref}
\usepackage[utf8]{inputenc}
\usepackage[T1]{fontenc}
\usepackage{lmodern}
\usepackage{fontawesome}
\usepackage{hyperref}
\usepackage{tikz}
\usepackage{subcaption}
\usepackage{mathtools}
\usepackage{xcolor}
\usepackage{amsmath,amssymb}
\usepackage{booktabs}

\usetikzlibrary{positioning}
\usetikzlibrary{automata,positioning,arrows,matrix,backgrounds,calc}
\usetikzlibrary{decorations.text}
\usetikzlibrary{decorations.pathmorphing}
\usetikzlibrary{shapes.geometric}
\usetikzlibrary{arrows.meta}

\usepackage{pgf}
\usepackage{relsize}

\newcommand{\simunet}{\texttt{SIMUnet}}
\newcommand{\colibri}{\texttt{Colibri}}
\newcommand{\fitmaker}{\texttt{fitmaker}}
\newcommand{\smefit}{\texttt{SMEFiT}}
\newcommand{\nnpdf}{\texttt{NNPDF4.0}}

\newcommand{\bea}{\begin{align}}
\newcommand{\eea}{\end{align}}

\usetikzlibrary{decorations,patterns,arrows.meta}
\usetikzlibrary{decorations.pathreplacing,decorations.markings}

\tikzset{
    ->-/.style={decoration={
  markings,
  mark=at position .65 with {\arrow{Latex}}},postaction={decorate}},
    -<-/.style={decoration={
  markings,
  mark=at position .6 with {\arrow{<}}},postaction={decorate}},
    ->/.style={decoration={
  markings,
  mark=at position .4 with {\arrow{>}}},postaction={decorate}},
}
\title{Unveiling the low- and high-energy frontier: \\
a simultaneous SMEFT-PDF fit with parity violation}

\author[a]{Ella Cole,}
\author[b,c]{Tom Tong,}
\author[a]{Maria Ubiali}

\affiliation[a]{DAMTP, University of Cambridge, Wilberforce Road, Cambridge CB3 0WA, UK }
\affiliation[b]{Center for Particle Physics Siegen, University Siegen, 57068 Siegen, Germany}
\affiliation[c]{Leibniz Institute for Astrophysics Potsdam (AIP), 14482 Potsdam, Germany}

\emailAdd{egc41@cam.ac.uk}

\abstract{
We present a comprehensive global analysis of Standard Model Effective Field Theory (SMEFT) operators utilising a diverse set of high- and low-energy experimental probes. By systematically matching the SMEFT to the Low-Energy Effective Field Theory (LEFT), we incorporate parity-violating observables into a global fit of 15 independent Wilson coefficients. 
This low-energy dataset, comprising parity-violation observables, neutron and nuclear $\beta$-decays, as well as leptonic and semileptonic kaon and pion decays, is combined with high-energy measurements including high-invariant-mass Drell-Yan data in both neutral- and charged-current channels, with the latter included for the first time, Higgs boson measurements, 
Electroweak Precision Observables (EWPOs) from LEP, and diboson production at LEP and the LHC. 
Crucially, this framework features a simultaneous fit of both the proton Parton Distribution Functions (PDFs) and the SMEFT Wilson coefficients using the public {\tt SIMUnet} framework. 
We quantify the direct interplay between SMEFT operators and hadron structure, demonstrating the robustness of the SMEFT bounds and the relative stability of 
the resulting PDFs with respect to the inclusion of SMEFT effects, with only a moderate increase in their uncertainties.}
\begin{document}
\maketitle

\section{Introduction}
\label{sec:intro}

The Standard Model (SM) of particle physics remains an extraordinarily successful description of fundamental interactions, yet it is widely 
accepted as an effective low-energy manifestation of a more fundamental theory. In the absence of direct discoveries of new particles 
at the Large Hadron Collider (LHC), the Standard Model Effective Field Theory (SMEFT) has emerged as the most popular model-independent framework to 
systematically search for subtle deviations from SM predictions. By parametrising heavy New Physics via a systematic expansion of higher-dimensional 
operators, SMEFT allows us to connect disparate experimental sectors under a unified theoretical framework~\cite{Brivio:2017vri,Isidori:2023pyp,Falkowski:2023hsg}.

In the SMEFT framework, short-distance dynamics is encoded in a set of higher-dimensional operators, whose Wilson coefficients induce correlated shifts across 
various sets of observables. This common language makes it possible to combine data that are experimentally and theoretically very different, 
while retaining the correlations implied by the structure of the SMEFT contributions.
Global SMEFT interpretations are particularly powerful because individual measurements often constrain only certain linear combinations of Wilson coefficients. 
Combining many sectors can therefore lift degeneracies and identify directions in parameter space that are weakly constrained by any single class of observables; 
see Refs.~\cite{Ellis:2020unq,Bartocci:2023nvp,deBlas:2025vqz,ATLAS:2026fyh} for recent examples of global SMEFT analyses. 

Recent investigations have shown that in the High-Luminosity phase of the LHC (HL-LHC) there may be a significant interplay between SMEFT effects and 
the parton distribution functions (PDFs) used to predict the SM contribution~\cite{Carrazza:2019sec,Greljo:2021kvv,Kassabov:2023hbm,Hammou:2023heg,Hammou:2024xuj,Cole:2026eex}, 
and that SMEFT effects may be hidden away in the large-$x$ PDFs. 
This is especially relevant in the high-energy tails of Drell-Yan distributions, where contact interactions and modified quark-antiquark luminosities can produce similar 
distortions. Several strategies can be applied to address such an issue~\cite{Hammou:2023heg,Hammou:2024xuj,Cole:2026eex}, with simultaneous PDF and SMEFT fits being the preferred option when viable, 
as they are able to disentangle the SMEFT from the PDF contribution and maximally constrain both the SMEFT and PDF sectors.
When a full simultaneous fit is not feasible, an alternative and computationally cheaper approach is offered by semiparametric methods, pioneered for LHC phenomenology in Ref.~\cite{Zhu:2026iqj}, which allow the systematic evaluation of a correction factor to the SMEFT coefficients arising from the PDF nuisance parameters.
The first simultaneous extraction of PDFs and a SMEFT Wilson coefficient from LHC data was performed by CMS using inclusive jet cross sections to constrain a four-quark contact interaction~\cite{CMS:2021yzl}. More recently, the {\tt xFitter} framework has been extended to perform simultaneous fits of PDFs with $\alpha_s$, $m_t$, and SMEFT Wilson coefficients~\cite{Shen:2024sci,MendizabalMorentin:2026sws}, and further to include the effective weak mixing angle~\cite{MendizabalMorentin:2026sws}.
The open-source \simunet{} framework was developed exactly to address this problem by fitting PDFs and SMEFT Wilson coefficients 
simultaneously within the \nnpdf{} methodology~\cite{Iranipour:2022iak,NNPDF:2021njg}. 
More recently, the \simunet{} dataset has been extended beyond the original Drell-Yan applications~\cite{Iranipour:2022iak} 
to include EWPOs, Higgs, diboson, and other collider observables, such as top-quark sector observables that are sensitive 
to dimension-six operators~\cite{Kassabov:2023hbm,Costantini:2024xae}. The \colibri{} framework~\cite{Costantini:2025agd} extends 
these simultaneous fitting capabilities to generic PDF parametrisations~\cite{Costantini:2025wxp}.  

Low-energy precision observables add an essential and largely PDF-independent component to this programme. Nuclear and neutron $\beta$-decays, 
pion and kaon decays, parity-violating electron scattering (PVES), atomic parity violation (APV), parity-violating deep-inelastic scattering (PVDIS), and coherent elastic 
neutrino--nucleus scattering (CE$\nu$NS) probe charged- and neutral-current interactions. These measurements are sensitive to 
semileptonic four-fermion operators and to modified electroweak couplings, and can therefore constrain directions that are only weakly resolved by collider 
data alone. They also play a central role in discussions of first-row CKM unitarity and related low-energy anomalies, where the interpretation of 
precision weak data can depend on the simultaneous treatment of EWPOs and LHC or future FCC-ee measurements~\cite{Cirigliano:2022qdm,Cirigliano:2023nol,Crivellin:2021bkd,Gialidi:2026tao}. 
The public {\tt smelli} framework~\cite{Aebischer:2018iyb,Straub:2018kue,Smolkovic:2026cba}, which built on {\tt flavio}, provides a 
global SMEFT likelihood encompassing quark and lepton flavour physics, low-energy parity violation, electroweak precision tests, and other low-energy precision observables; its scope was subsequently extended to include a combined Higgs and electroweak likelihood~\cite{Falkowski:2019hvp} and high-mass 
Drell-Yan observables~\cite{Greljo:2022jac,Greljo:2023bab}. In contrast to {\tt smelli}, where PDFs enter as fixed external inputs, 
the present work fits PDFs and Wilson coefficients simultaneously, and extends the global analysis to include high-energy, PDF-sensitive observables, such as 
the full set of high-invariant-mass neutral and charged-current Drell-Yan distributions that provide crucial constraints on the large-$x$ quark and antiquark distributions 
in PDF fits and the whole set of top observables described in Refs.~\cite{Kassabov:2023hbm,Costantini:2024xae} that provide crucial constraints to the large-$x$ gluon distributions.

In this work, we perform a simultaneous global fit of PDFs and a 15-dimensional space of SMEFT Wilson coefficients.  
Our goal is to integrate low-energy observables and their SMEFT operator dependence into \simunet{}, and to reassess the resulting global constraints. 
Compared with the low-energy global analysis of Ref.~\cite{Cirigliano:2023nol}, which included neutron and nuclear $\beta$-decays together with pion and kaon decay observables, 
we expand the set of precision inputs by incorporating parity-violating low-energy measurements. 
At the same time, we complete the connection to the collider programme by combining these inputs with the existing \simunet{} high-energy dataset, including neutral- 
and charged-current high-mass Drell-Yan measurements. Crucially, in all previous \simunet{} publications, high-mass Drell-Yan datasets were analysed in isolation or within highly 
restricted sector combinations to demonstrate the proof of principle of PDF-SMEFT cross-talk. This work marks the first time 
that high-mass Drell-Yan tails are integrated into a full global analysis alongside EWPOs, Higgs, and diboson. 
The addition of the new low- and high-energy observables extends the previous global analysis and makes it possible to study, within a single framework, 
low-energy inputs together with the higher-energy measurements, 
and the PDF-dependent collider observables that enter the simultaneous fit. This enlarged dataset allows us to revisit the complementarity 
of low- and high-energy probes in a framework that consistently accounts for experimental information and SMEFT parameter correlations.
Moreover, by fitting the PDFs and the 15 SMEFT operators simultaneously, we isolate and assess the interplay between QCD structure and electroweak-scale 
effective interactions. We explicitly evaluate whether a joint fit relaxes or sharpens SMEFT bounds and determine the quantitative impact of potential 
New Physics effects on the determination of the PDFs.

The structure of the paper is as follows. In Section~\ref{sec:approach} we describe the fitting strategy and the implementation of the 
collider and low-energy datasets entering the analysis. 
The results of the fixed-PDF and simultaneous PDF--SMEFT fits are presented in Section~\ref{sec:results}. We conclude in Section~\ref{sec:conclusion}, 
while Appendix~\ref{app:datasets} collects further details on the low-energy observables used in the fits.
\section{Analysis settings}
\label{sec:approach}

In this section, we summarise the theoretical and experimental ingredients that enter our analysis. We begin by defining the 15 SMEFT operators considered in this work (Sect.~\ref{subsec:theory}). We then provide a comprehensive breakdown of the experimental datasets that enter our global SMEFT-only and simultaneous SMEFT-and-PDF fits (Sect.~\ref{subsec:data}). Finally, we outline the core components of our simultaneous fitting methodology (Sect.~\ref{subsec:methodology}).

\subsection{SMEFT operators}
\label{subsec:theory}

We adopt the same flavour assumptions and operator-basis conventions as used in the \simunet{} framework, 
following the standard definitions of Refs.~\cite{Degrande:2020evl,Aguilar-Saavedra:2018ksv}. 
The flavour symmetry group is given by $U(3)_l \times U(3)_e \times U(3)_d \times U(2)_u \times U(2)_q$, i.e. the operators that contain
top quarks (right-handed $t$ and $SU(2)$ doublet $Q$) are singled out. Consistently we work in a five-flavour scheme in which the
only massive fermion in the theory is the top.
As far as the electroweak input scheme is concerned, we work in the $m_W$-scheme, meaning that the $4$
electroweak inputs are
$\{m_W, G_F, m_h, m_Z\}$.

For the low-energy observables, the matching between the Low-Energy Effective Field Theory (LEFT) and the SMEFT is performed 
following the analytical framework detailed in Appendix~A.2 of Ref.~\cite{Cirigliano:2023nol}. To incorporate these 
low-energy corrections into the \simunet{} framework, we start from the analytical expressions of Ref.~\cite{Cirigliano:2023nol}, 
normalise all observables such that the SM prediction equals unity, and scale both the SMEFT contributions and the experimental central values and 
uncertainties accordingly. We then project these expressions onto the above-mentioned \simunet{} flavour symmetry assumptions, 
perform the explicit basis transformation from the operator conventions of Ref.~\cite{Cirigliano:2023nol} to those of \simunet{}, 
and encode the resulting SMEFT weight dependencies into custom \simunet{} data files (SIMU-factors).

In total, we include 15 dimension-six operators from the SMEFT in our global analysis, spanning the Higgs, diboson, 
electroweak precision, and low-energy sectors.
\begin{figure}[thb!]
    \centering
    \includegraphics[width=0.5\textwidth]{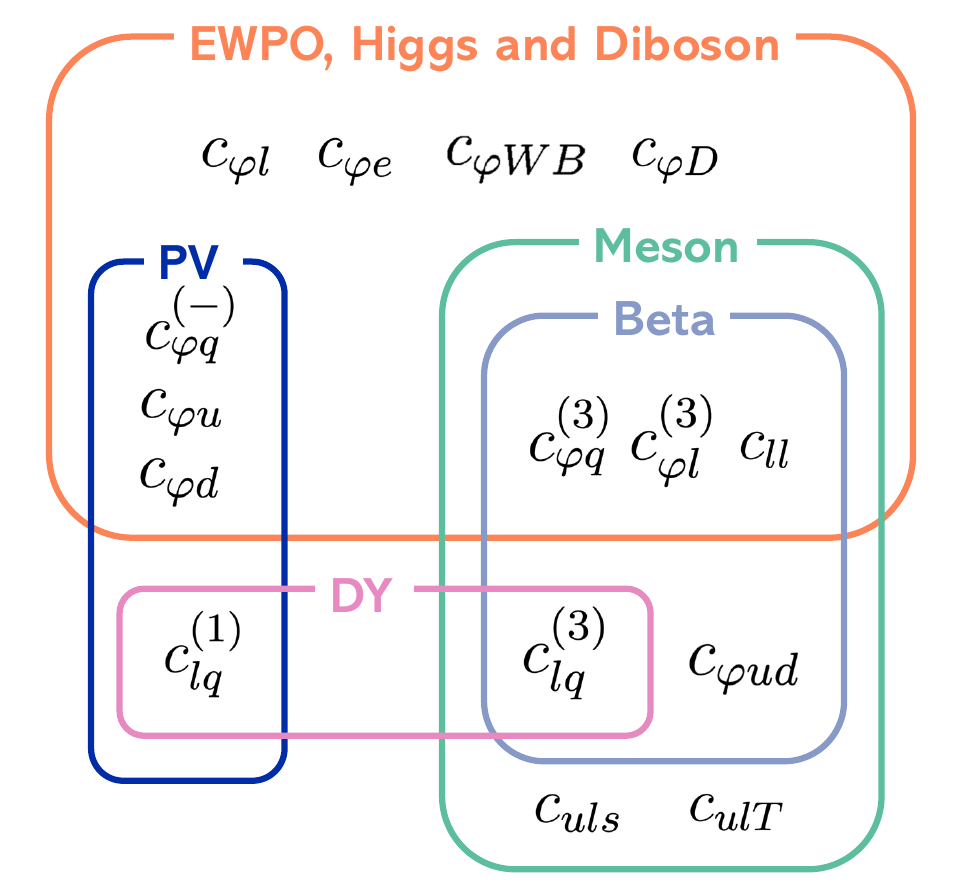}
    \caption{Schematic representation of the data categories entering this analysis and their overlapping dependences on the 15 dim-6 SMEFT operators, adapted from Ref.~\cite{Ellis:2020unq}. }
\label{fig:operators}
\end{figure}
In Fig.~\ref{fig:operators}, we display the targeted SMEFT WCs, adhering to the name conventions established in 
Refs.~\cite{Kassabov:2023hbm,Ethier:2021bye}. The schematic map, adapted from Ref.~\cite{Ellis:2020unq}, classifies the specific 
experimental sectors sensitive to each WC, illustrating the tightly interconnected nature of a truly global SMEFT determination. 
The overlaps between the different regions show explicitly how a single operator simultaneously modifies multiple data categories. 
This schematic illustrates the interconnected web linking high-energy probes (EWPO, Higgs, diboson, and Drell-Yan) with low-energy measurements (parity violation, $\beta$-decays, and meson sectors) through their shared dependence on SMEFT operators.
For the specific set of parity violation observables included here, we consider only the operator combinations involving quark currents, since these are the combinations directly probed by our datasets.

\subsection{Datasets}
\label{subsec:data}

The simultaneous global analysis of SMEFT Wilson coefficients and proton PDFs incorporates both SMEFT-sensitive observables and baseline 
datasets that carry no New Physics dependencies. The complete dataset configuration is summarised in Table~\ref{tab:datasummary}, 
which categorises each sector by its total number of data points ($N_{\rm data}$) and its structural sensitivity to both SMEFT variations and PDF uncertainties.

The ``Other'' category encompasses the essential baseline datasets that do not carry explicit dependence on our chosen 15 SMEFT operators but are 
crucial for constraining the underlying PDFs. This background dataset includes deep-inelastic scattering (DIS) from HERA and fixed-target 
experiments, inclusive jet and di-jet production, and isolated photon data. 

Notably, we choose to exclude the top-quark sector from this analysis. While top-quark observables are highly sensitive to SMEFT effects, 
their inclusion introduces a large number of additional top-specific four-fermion and dipole operators. To isolate the clean interplay between 
the 15 electroweak and low-energy operators of interest and to avoid unconstrained flat directions from an overly inflated operator space, we defer 
the full integration of the top sector to future work, building on the standalone top-quark treatments in Ref.~\cite{Kassabov:2023hbm}. Additional 
information on all datasets included in the analysis is provided in Appendix~\ref{app:datasets}.

\begin{table}[ht]
\centering
\begin{tabular}{l | c | c | c}
Data type & $N_{\rm data}$ & SMEFT-dependent & PDF-dependent \\
\hline
High-mass Drell-Yan   & 259  & Yes  & Yes \\
EWPO                  & 44   & Yes  & No    \\
Higgs                 & 73   & Yes  & No$^*$    \\
Diboson               & 82   & Yes  & No$^*$ \\
$\beta$-decay         & 42   & Yes  & No    \\
Mesons                & 10   & Yes  & No    \\
PV observables        &  6   & Yes  & No     \\
Other (DIS, Jets, etc.) & 4364 & No   & Yes \\
\hline
Total                 & 4880 &      &     \\
\hline
    \end{tabular}
    \caption{Summary of the experimental datasets included in the global analysis, indicating their sensitivity to SMEFT operators entering our analysis and PDFs. 
    *The Higgs signal strengths and STXS measurements have a mild PDF dependence that is neglected here, given their strong sensitivity to SMEFT corrections. 
    For diboson, the LEP measurements are PDF-independent by definition; the LHC diboson measurements are technically PDF-dependent, but their sensitivity 
    is negligible compared to competing constraints in the same Bjorken-$x$ range from single-boson production, and are therefore treated as PDF-independent.}
    \label{tab:datasummary}
\end{table}

The dataset explored in this study significantly expands upon the frameworks established in 
Refs.~\cite{Greljo:2021kvv,Iranipour:2022iak,Kassabov:2023hbm,Costantini:2024xae}. These preceding analyses grew out of the core \nnpdf{} 
NNLO global methodology~\cite{NNPDF:2021njg}, which comprises $N_{\rm data} = 4618$ data points spanning deep-inelastic lepton-proton scattering 
(HERA, NMC, SLAC, BCDMS, CHORUS, and NuTeV), fixed-target Drell-Yan measurements (Fermilab E605 and E866), Tevatron $p\bar{p}$ collisions, and an 
extensive suite of standard LHC Run I and Run II measurements. 

To maximise sensitivity to potential electroweak modifications, the high-mass neutral-current Drell-Yan sector from the standard \nnpdf{} baseline 
was previously augmented by CMS measurements at $\sqrt{s}=8$ and $13$\,TeV, adding 65 high-energy data points~\cite{Greljo:2021kvv,Iranipour:2022iak}. 
In Ref.~\cite{Costantini:2024xae}, this was further extended to incorporate EWPOs from LEP, Higgs boson signal strengths, and LHC diboson production. 
Tables~\ref{tab:dataset_EWPO}, \ref{tab:dataset_Higgs}, and \ref{tab:dataset_diboson} in Appendix~\ref{app:datasets} outline the specifics 
(e.g., centre-of-mass energies, integrated luminosities, and journal references) for these sectors. Crucially, the theoretical predictions for the 
Higgs and diboson sectors are evaluated using a frozen PDF set, given that the direct impact of these specific datasets on the extraction of proton 
structure is negligible compared to their experimental precision.

In this work, we introduce two major expansions to this high-energy framework. First, alongside the neutral-current channels, we include, for the first time in a SMEFT fit, the ATLAS high-mass charged-current Drell-Yan measurement at $\sqrt{s}=13$\,TeV~\cite{ATLAS:2025hhn}. This is the first experimental measurement of high-transverse-mass charged-current Drell-Yan observables, previously only explored via HL-LHC projections, adding 12 highly sensitive high-mass data points to the Drell-Yan sector. Second, we integrate a dedicated selection of low-energy precision probes building upon 
Ref.~\cite{Cirigliano:2023nol}, supplemented uniquely by parity-violating observables (for more details see Appendix~\ref{app:datasets}):

\begin{itemize}
    \item \textbf{$\beta$-decays:} We include the free neutron lifetime, $0^+\to 0^+$ superallowed nuclear transitions, and mirror transitions. 
    Beyond decay rates, we incorporate key angular correlation coefficients, such as the neutrino-electron asymmetry parameter $a$ and the $\beta$-asymmetry parameter $A$.
    \item \textbf{Meson decays:} We incorporate two-body leptonic decays via the absolute decay rate $\Gamma(K\to \mu\nu_\mu)$ and the precise ratios 
    $R_\pi=\Gamma(\pi\to e\nu_e)/\Gamma(\pi\to \mu\nu_\mu)$, $R_K=\Gamma(K\to e\nu_e)/\Gamma(K\to \mu\nu_\mu)$, and $R_\mu=\Gamma(K\to \mu\nu_\mu)/\Gamma(\pi\to \mu\nu_\mu)$ (Table~\ref{tab:mesons}). Three-body semileptonic decays ($K_{l3}$ and pion $\beta$-decay) and their differential spectral shapes are also included to bound scalar and 
    tensor operators.
    \item \textbf{Parity-violating (PV) observables:} We incorporate parity-violating electron scattering (PVES via $Q_\text{w}^p$ at JLab), atomic parity violation (APV in $^{133}\text{Cs}$), parity-violating deep-inelastic scattering (PVDIS at JLab), and coherent elastic neutrino-nucleus scattering (CE$\nu$NS from the COHERENT CsI and liquid Argon targets). 
    We derive the \simunet{} K-factors for the PV observables using NNPDF23LO, following the analysis of Ref.~\cite{Crivellin:2021bkd}, although we do not expect them to have a significant impact on the PDFs given the large number of constraints we have in our PDF analysis and the kinematic cuts in $W^2$ that exclude $x$ values close to 1. It is worth noting that a recent study~\cite{Whitehill:2026swa} highlighted the interplay between PVDIS observables and PDFs. Assessing the extent to which this interplay affects our K-factors would be interesting, and we leave this investigation to future work.
\end{itemize}

The highly flexible architecture of \simunet{} allows us to cleanly handle both PDF-dependent and PDF-independent data streams. 
For instance, we can include observables like the fine-structure constant $\alpha_e$ or Higgs Simplified Template Cross Sections (STXS)—which are 
completely decoupled from or only mildly affected by hadron structure—while fully accounting for their heavy SMEFT parameter dependence.

The simultaneous fit includes 4880 data points, of which 516 are sensitive to the 15 SMEFT operators considered here. Of these, 257 are treated as PDF-independent. These comprise EWPOs, Higgs and diboson measurements, and the low-energy observables. The pure PDF fit therefore uses $4880-257=4623$ points, while the SMEFT-only fit uses up to 516 points.

\subsection{Methodology}
\label{subsec:methodology}

Our strategy is to exploit the simultaneous fitting capabilities of the \simunet{} framework to systematically dissect the 
interplay between high-energy collider observables, low-energy precision measurements, and proton structure. 

The core of the \simunet{} methodology relies on a neural-network parametrisation of the PDFs integrated into a Monte Carlo replica 
framework, optimised concurrently with a linear expansion of the SMEFT corrections via $\chi^2$ minimisation.
For the low-energy charged-current observables we use effective uncertainties calibrated to the Standard Model fit of Ref.~\cite{Cirigliano:2023nol}. This allows us to keep the low-energy parameters fixed in the simultaneous fit. Appendix~\ref{app:low-energy-data} describes the calibration.

To isolate the impact of each dataset sector, we perform our fits in sequential phases. 
First, we perform a series of fixed-PDF SMEFT fits where the low-energy datasets and the high-mass Drell-Yan measurements 
are added one by one. This reveals how low-energy inputs can structurally break degeneracies in the Wilson coefficient space. 
Second, we transition to the full simultaneous \simunet{} configuration, fitting the 15 WCs and the PDFs concurrently. 
This final step allows us to quantify exactly how the newly added parity-violating and low-energy observables modify 
the joint PDF-SMEFT bounds, ensuring that any hints of New Physics in the high-energy tails are not mistakenly absorbed 
into or obscured by the proton's sea quark distributions.
\section{Results}
\label{sec:results}

In this section, we present the numerical results of our global analyses. We explore the performance of our 
framework across different dataset configurations to systematically isolate the impact of the newly incorporated low-energy 
precision measurements and the high-mass charged-current Drell-Yan data. In Sect.~\ref{subsec:fixed_smeft}, we analyse these effects 
within a fixed-PDF framework, evaluating how new datasets alter the individual bounds, modify two-dimensional SMEFT WC correlations, 
and characterise the relative constraining power of each sector via a Fisher information analysis. In Sect.~\ref{subsec:PVeffect} we 
single out the effect of Parity-Violating observables. Subsequently, in 
Sect.~\ref{sec4.2:fixed_vs_simu}, we lift the assumption of keeping the PDF fixed and present the results of the full simultaneous 
fit of the PDFs and the 15 SMEFT Wilson coefficients, quantifying the direct physical interplay between possible New Physics effects and 
proton structure.

\subsection{The global SMEFT analysis within a fixed-PDF framework}
\label{subsec:fixed_smeft}

The specific effect of the parity-violating observables on the SMEFT bounds is discussed separately in Sect.~\ref{subsec:PVeffect}.

To isolate the bounding power of our expanded dataset without the simultaneous effects of PDF variations, 
we perform a sequence of SMEFT-only fits where the underlying parton distributions are kept frozen to the baseline PDF set, 
which includes only SMEFT-independent datasets, i.e. datasets for which SMEFT corrections are negligible compared to the experimental uncertainties\footnote{To this category belong the DIS data, as well as the Drell-Yan data included in 
the NNPDF4.0 analysis~\cite{NNPDF:2021njg}, which have large statistical uncertainties in the large invariant mass bins.}.
This allows us to track the sequential tightening of the parameter space as new experimental information is activated.

Our SMEFT analysis includes up to 516 data points sensitive to SMEFT corrections, 
and the data points belong to the various sectors listed in Table~\ref{tab:datasummary}, comprising EWPOs, 
Higgs boson signal strengths, neutral-current and charged current Drell-Yan (DY), diboson production channels, 
neutron and nuclear $\beta$-decays, as well as parity violating observables. We construct four distinct fit scenarios:
\begin{enumerate}
    \item[(i)] \textbf{Baseline Collider Fit:} Utilising only the core collider data (Drell-Yan, Higgs, EWPOs, diboson) 
    with the exclusion of the new high-mass charged-current Drell-Yan ATLAS measurement of Ref.~\cite{ATLAS:2025hhn} comprising 12 data points, 
    which we want to include separately to display the constraining power of such measurement ($N_{\rm data} = 446$).
    \item[(ii)] \textbf{Collider + Low-Energy Fit:} Adding the neutron, mesons and nuclear $\beta$-decay datasets ($N_{\rm data} = 498$).
    \item[(iii)] \textbf{Collider + Low-Energy + PV Fit:} Adding Parity Violation observables ($N_{\rm data} = 504$).
    \item[(iv)] \textbf{Collider + Low-Energy + PV + HMCC DY Fit (Global SMEFT Fit):} Incorporating both the low-energy precision data and the newly released ATLAS high-mass charged-current Drell-Yan (CCDY) data of Ref.~\cite{ATLAS:2025hhn} ($N_{\rm data} = 516$).
\end{enumerate}

\begin{figure}[ht!]
    \centering
    \includegraphics[width=0.8\linewidth]{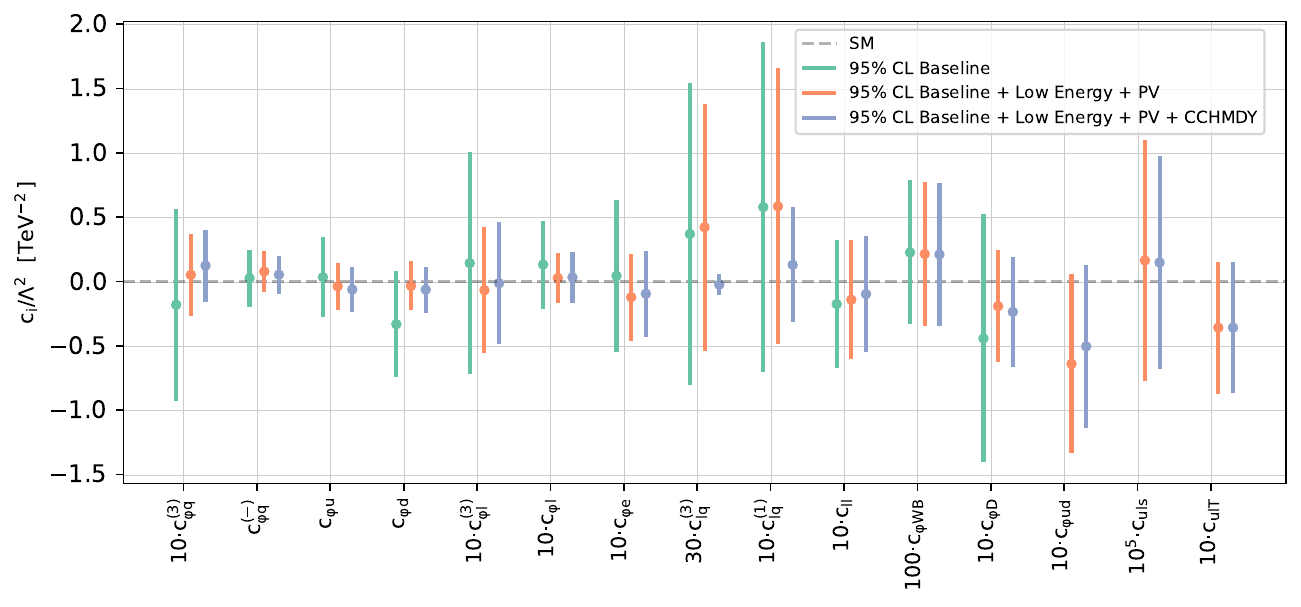}
    \caption{The 95\% confidence intervals for the 15 SMEFT Wilson coefficients extracted from the fixed-PDF fits before the inclusion of low-energy data (green), after the inclusion of low-energy observables (orange), and with the final addition of the ATLAS high-mass charged-current Drell-Yan dataset (blue).}
    \label{fig:fixed_pdf_bounds}
\end{figure}

The marginalised 95\% confidence intervals for the 15 Wilson coefficients across these stages (i), (ii), and (iv) are illustrated 
in Fig.~\ref{fig:fixed_pdf_bounds}. We will focus on the effect of the PV observables, scenario (iii), in Sect.~\ref{subsec:PVeffect}.   
We obtain these bounds by fitting the Wilson coefficients independently to each Monte Carlo pseudodata replica using \simunet{}~\cite{Iranipour:2022iak}. The replicas account for experimental and PDF uncertainties and use the effective uncertainties of Appendix~\ref{app:low-energy-data} for the low-energy charged-current data. 

\begin{figure}
    \centering
    \includegraphics[width=1\linewidth]{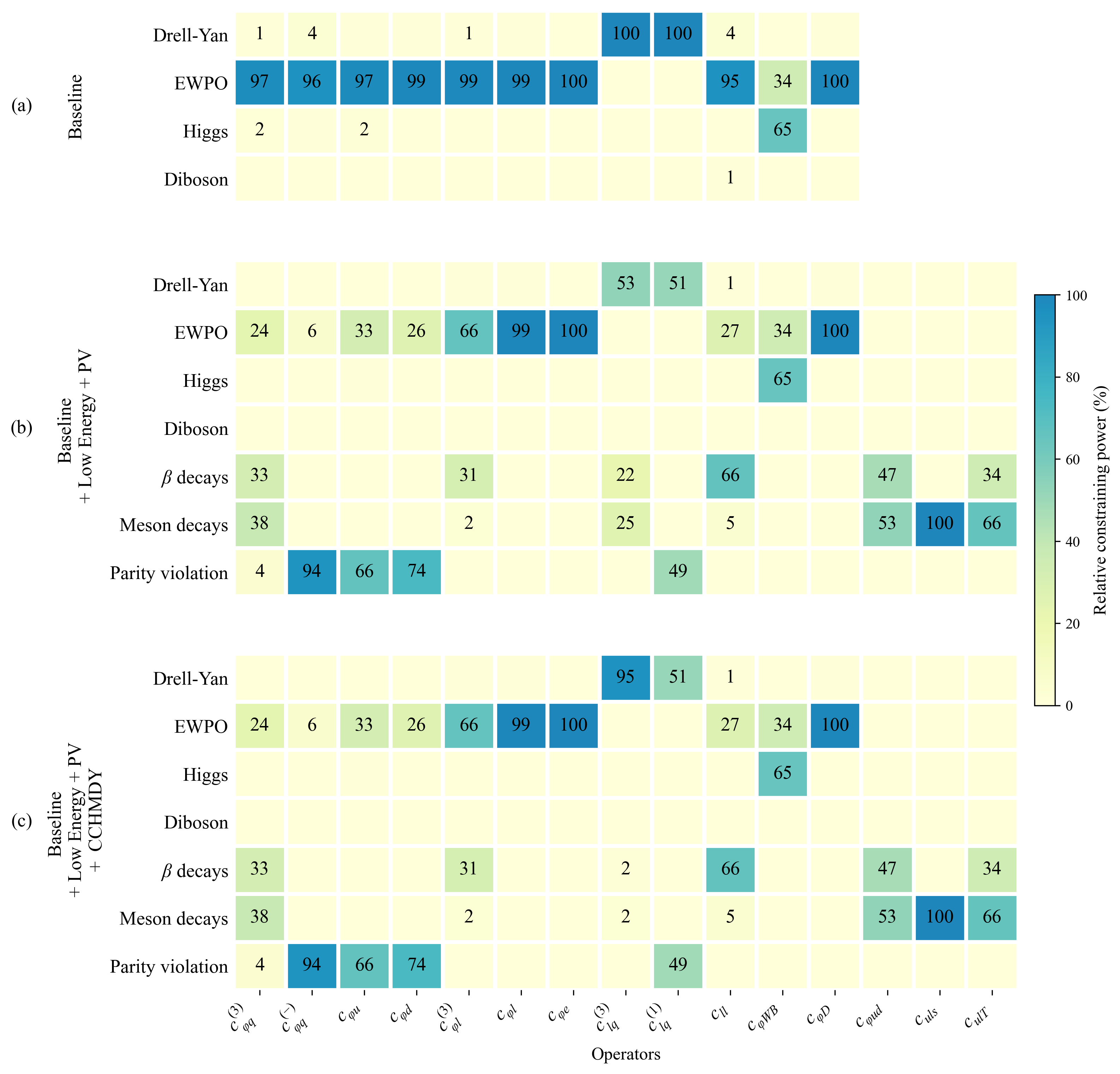}
    \caption{Relative constraining power (expressed in \%) of each experimental sector on the individual SMEFT operators for the fixed-PDF fits using (a) the baseline collider datasets \textbf{(i)}, (b) the baseline + low energy + parity violation observables \textbf{(iii)}, and (c) the full dataset with the CCHMDY dataset included \textbf{(iv)}.}
    \label{fig:fisher}
\end{figure}

\begin{figure}[ht!]
    \centering
    \includegraphics[width=0.5\linewidth]{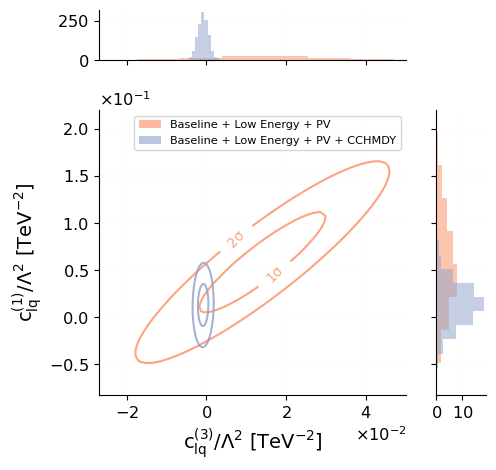}
    \caption{Two-dimensional confidence regions (1$\sigma$ and 2$\sigma$) in the $(c_{lq}^{(1)}, c_{lq}^{(3)})$ plane, comparing the fit before (orange) and after (blue) the inclusion of the ATLAS high-mass charged-current Drell-Yan dataset, illustrating the lifting of a prominent flat direction.}
    \label{fig:flat_dir_ellipse}
\end{figure}

The primary takeaway from Fig.~\ref{fig:fixed_pdf_bounds} is the dramatic impact of the low-energy precision data in shrinking the 
allowed parameter space for several previously poorly constrained directions. While high-energy collider data provide excellent 
sensitivity to operators that grow quadratically with the centre-of-mass energy (such as four-fermion contact interactions in the Drell-Yan tails), 
they frequently suffer from blind spots or flat directions when multiple vertex corrections and four-fermion operators are turned on simultaneously. 
The low-energy precision inputs act as an exceptional anchor, cleanly isolating individual operators at low momentum transfer where 
competing high-energy operator dependencies are absent. 
The second takeaway from Fig.~\ref{fig:fixed_pdf_bounds} is the particularly strong impact of the ATLAS high-mass charged-current
Drell-Yan measurement of Ref.~\cite{ATLAS:2025hhn} on the triplet four-fermion coefficient $c_{lq}^{(3)}$.
Before this dataset is included, $c_{lq}^{(3)}$ is constrained substantially by the $\beta$-decay and meson-decay sectors, as reflected by the Fisher decomposition
in Fig.~\ref{fig:fisher}b. However, in the absence of the charged-current high-mass tail, these constraints still leave an almost flat direction
in the $(c_{lq}^{(1)}, c_{lq}^{(3)})$ plane, visible in Fig.~\ref{fig:flat_dir_ellipse}. The charged-current Drell-Yan process probes the triplet interaction
with a complementary high-energy lever arm, thereby lifting this degeneracy and sharply increasing the relative constraining power on $c_{lq}^{(3)}$,
as shown by the comparison with the full Fisher profile in Fig.~\ref{fig:fisher}c.

\begin{figure}[ht!]
    \centering
    \begin{subfigure}[b]{0.45\textwidth}
        \centering
        \includegraphics[width=\textwidth]{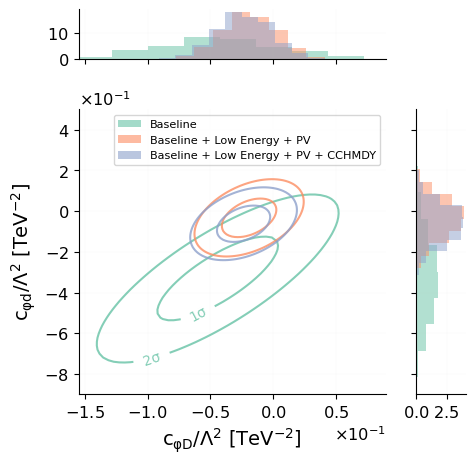} 
        \caption{}
        \label{fig:ellipse_a}
    \end{subfigure}
    \hfill
    \begin{subfigure}[b]{0.45\textwidth}
        \centering
        \includegraphics[width=\textwidth]{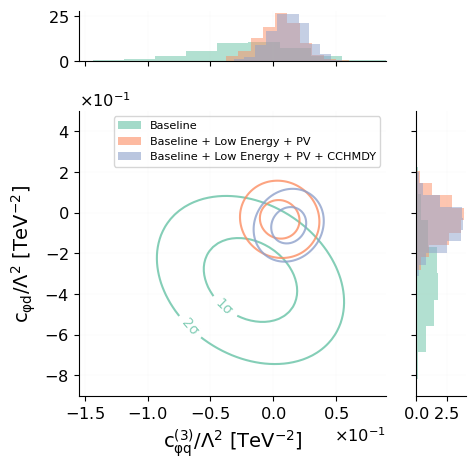} 
        \caption{}
        \label{fig:ellipse_b}
    \end{subfigure}
    \caption{Two-dimensional confidence contours (1$\sigma$ and 2$\sigma$) for the coefficient pairs $(c_{\varphi d}, c_{\varphi D})$ (a) and $(c_{\varphi d}, c_{\varphi q}^{(3)})$ (b) within the fixed-PDF framework discussed in Sect.~\ref{subsec:fixed_smeft}. The green contours denote the baseline fit, the orange contours demonstrate the impact of adding low-energy inputs, and the blue contours illustrate the final constraints after adding the ATLAS CCDY data.}
    \label{fig:fixed_pdf_ellipse}
\end{figure}

This complementary behavior is visually apparent in the two-dimensional correlation contours displayed in Fig.~\ref{fig:fixed_pdf_ellipse}. 
For instance, in the $(c_{\varphi d}, c_{\varphi D})$ plane (Fig.~\ref{fig:ellipse_a}), the baseline high-energy fit exhibits an extended, 
tilted error ellipse representing a significant parameter degeneracy. The inclusion of the low-energy dataset (moving from the green 
to the orange contours) cleanly truncates this ellipse along its weakly constrained major axis, drastically reducing the overall volume 
of the allowed parameter space. Similarly, the $(c_{\varphi d}, c_{\varphi q}^{(3)})$ plane (Fig.~\ref{fig:ellipse_b}) demonstrates a powerful 
structural contraction upon the introduction of the low-energy precision measurements and a further improvement in the $c_{\varphi q}^{(3)}$ direction associated to 
the inclusion of the high-mass CC observable.

To rigorously quantify how the individual data sectors drive these improvements, we compute the relative profile of the Fisher information matrix. 
For a space of $N$ Wilson coefficients, the $N \times N$ Fisher matrix $F$ is defined as:
\begin{equation}
    F_{ij} = K^{i, T} \Sigma^{-1} K^j
\end{equation}
where $K^i$ represents the vector of linear SMEFT sensitivities for the $i$-th Wilson coefficient across the targeted observables, and $\Sigma$ 
is the covariance matrix used in the fit, including the effective uncertainties assigned to the charged-current decay data. 
The normalised, percentage-based relative constraining  power of a specific dataset sector $D$ on a given coefficient $c_i$ is then isolated via:
\begin{equation}
    \text{Relative constraining power of $D$ on } c_i = \frac{F_{ii}(D)}{\sum_{D'} F_{ii}(D')}
\end{equation}

The resulting sector-by-sector sensitivity profiles are presented in Fig.~\ref{fig:fisher} for the fit scenarios \textbf{(i)}, \textbf{(iii)}, and \textbf{(iv)}.
These diagrams reveal the physical origin of the constraints in our global SMEFT fit:

\begin{itemize}
    \item \textbf{The $\beta$-decay sector:} This sector emerges as the dominant source of information for operators modifying charged-current semileptonic 
    vertex structures and four-fermion operators,  strengthening the 
    bounds given by the EWPO on $c_{\varphi l}^{(3)}$ and $c_{\varphi q}^{(3)}$. Furthermore, 
    $\beta$-decays provide the single largest constraint on the purely leptonic operator $c_{ll}$. This dominance is driven by the structural matching condition to 
    LEFT below the electroweak scale, where modifications to the muon decay rate propagate into a shift in the extracted Fermi constant, expressed 
    as $\Delta G_F = (2 c_{\varphi l}^{(3)} - c_{ll}) / \sqrt{2} v^2$~\cite{Cirigliano:2023nol}. The $\beta$-decay sector also strongly constrains 
    the operator $c_{\varphi ud}$, which directly parameterises right-handed charged currents that are highly suppressed within the SM.
    \item \textbf{The Meson-Decay Sector:} As expected from leptonic and semileptonic pion and kaon decays, this sector heavily dominates 
    the constraints on the scalar and tensor semileptonic four-fermion operators, represented in our basis by $c_{u l s}$ and $c_{u l T}$. 
    The sector strengthen the constraints on the $c_{\varphi ud}$ Wilson coefficient and contributes alongside the $\beta$-decay sector and the EWPOs to constrain 
    the $c_{\varphi q}^{(3)}$ Wilson coefficient.
    \item \textbf{Parity-Violating (PV) Observables:} Because low-energy PVES, APV, and PVDIS processes are mediated by neutral-current interactions 
    that isolate parity-odd vector-axial ($V \times A$) electron-quark combinations, they provide high-precision constraints that are highly orthogonal 
    to the standard EWPO datasets. As a result, they supply critical constraining power for the four-fermion operator $c_{lq}^{(1)}$ and the Higgs-quark 
    vertex operators $c_{\varphi u}$, $c_{\varphi d}$, and $c_{\varphi q}^{(-)}$.
\end{itemize}

A key highlight of our analysis is the profound impact of the high-mass charged-current Drell-Yan (CCDY) data from ATLAS. Comparing the Fisher profiles in Fig.~\ref{fig:fisher}(b) and Fig.~\ref{fig:fisher}(c), we observe that the inclusion of the CC high-mass Drell-Yan dataset triggers a dramatic redistribution of constraining power for the triplet operator $c_{lq}^{(3)}$, with the Drell-Yan sector's share increasing from 53\% to 95\%.

The physical consequence of this energy-scale synergy is explicitly captured in Fig.~\ref{fig:flat_dir_ellipse}, which displays the joint replica distribution for the $(c_{lq}^{(1)}, c_{lq}^{(3)})$ four-fermion coefficients. Prior to including the CCDY data (orange ellipse), the fit exhibits a severe parameter degeneracy. This flat direction occurs because neutral-current high-mass Drell-Yan data are sensitive to a linear combination of both the singlet $O_{lq}^{(1)}$ and triplet $O_{lq}^{(3)}$ operators, making it impossible for neutral-current probes alone to un-correlate them cleanly. Because the charged-current Drell-Yan process ($u\bar{d} \to l^+\nu$) is mediated strictly by $W^{\pm}$ exchange, it is completely blind to the singlet operator $O_{lq}^{(1)}$ while remaining highly sensitive to the triplet operator $O_{lq}^{(3)}$. The addition of the ATLAS CCDY data (blue ellipse) thus completely lifts this degeneracy, collapsing the flat direction and significantly shrinking the independent uncertainties on both coefficients. This successful decorrelation represents a substantial step forward compared to earlier global SMEFT analyses (such as those in Ref.~\cite{Giani:2023gfq}), which lacked both the high-precision low-energy inputs and the high-mass charged-current tails needed to break these specific four-fermion degeneracies.

\subsection{The effect of Parity Violation observables}
\label{subsec:PVeffect}

In this section, we isolate the specific performance of the newly added parity-violating (PV) observables within this simultaneous framework, 
corresponding to scenario (iii) of the sequential dataset inclusion described in Sect.~\ref{subsec:fixed_smeft}. 
\begin{figure}[ht!]
    \centering
    \includegraphics[width=0.5\linewidth]{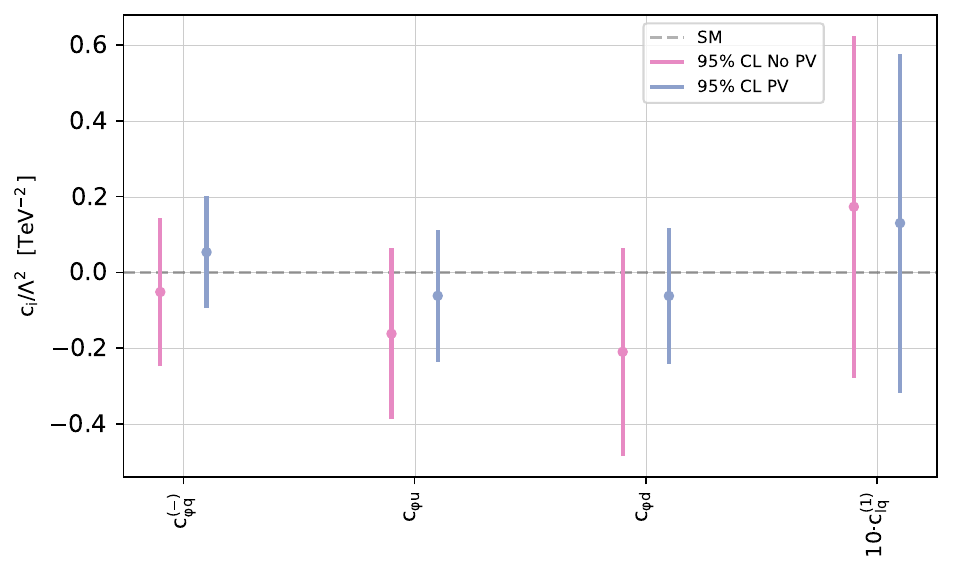}
    \caption{The 95\% confidence intervals for the Wilson coefficients sensitive to neutral-current interactions, comparing the global fit before (pink) and after (purple) the specific inclusion of the parity-violating datasets.}
    \label{fig:pv_bounds_impact}
\end{figure}

Fig.~\ref{fig:pv_bounds_impact} details the direct individual tightening of the 95\% C.L. intervals for the subset of operators sensitive to neutral-current quark couplings. 
The inclusion of the PV observables (moving from pink to purple bars) induces an unmistakable narrowing of the bounds, particularly for the right-handed Higgs-quark 
vertex corrections $c_{\varphi u}$ and $c_{\varphi d}$. 

\begin{figure}[ht!]
    \centering
    \begin{subfigure}[b]{0.45\textwidth}
        \centering
        \includegraphics[width=\textwidth]{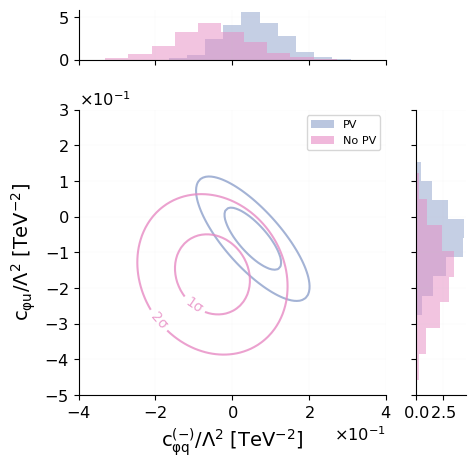} 
        \caption{}
        \label{fig:pv_ellipse_a}
    \end{subfigure}
    \hfill
    \begin{subfigure}[b]{0.45\textwidth}
        \centering
        \includegraphics[width=\textwidth]{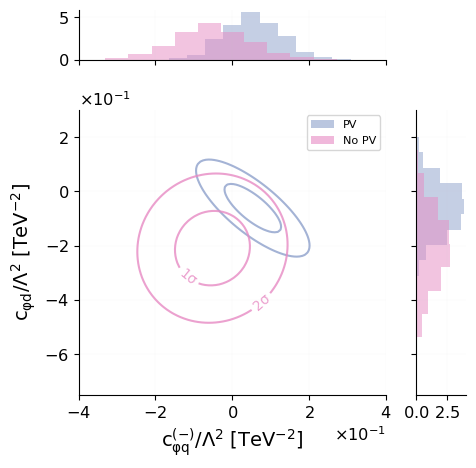} 
        \caption{}
        \label{fig:pv_ellipse_b}
    \end{subfigure}
    \caption{Two-dimensional confidence contours showing the distribution of the best-fit replicas for the operator pairs $(c_{\varphi u}, c_{\varphi q}^{(-)})$ (a) and $(c_{\varphi d}, c_{\varphi q}^{(-)})$ (b). The pink contours correspond to the fit excluding parity-violating data, while the blue contours demonstrate the tightened constraints after the inclusion of the PV datasets.}
    \label{fig:pv_ellipses}
\end{figure}
This targeted impact is further deconstructed in the two-dimensional replica distributions shown in Fig.~\ref{fig:pv_ellipses}. In both the $(c_{\varphi u}, c_{\varphi q}^{(-)})$ plane (Fig.~\ref{fig:pv_ellipse_a}) and the $(c_{\varphi d}, c_{\varphi q}^{(-)})$ plane (Fig.~\ref{fig:pv_ellipse_b}), the inclusion of the PV dataset contracts the allowed regions, thereby improving the constraints on the corresponding SMEFT coefficients. The accompanying changes in the orientation and shape of the contours indicate that the PV measurements probe combinations of coefficients that differ from those constrained by the remaining datasets.

\subsection{Interplay between SMEFT and PDFs in the Simultaneous Fit}
\label{sec4.2:fixed_vs_simu}

Having thoroughly mapped out the fixed-PDF landscape, we now lift the assumption of fixed PDFs to 
perform the full simultaneous fit of the 15 SMEFT Wilson coefficients and the proton PDFs via the joint \simunet{} optimization framework.

\begin{figure}[h]
    \centering
    \includegraphics[width=0.8\linewidth]{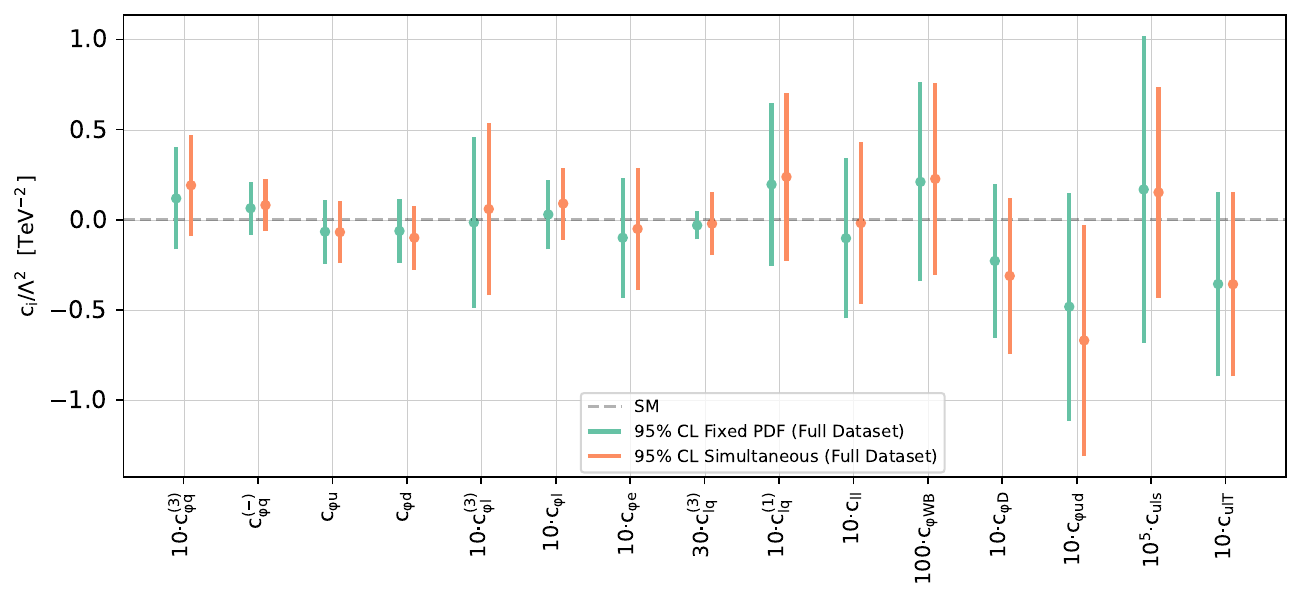}
    \caption{Comparison of the 95\% confidence bounds on the 15 SMEFT Wilson coefficients extracted from the fixed-PDF global fit (green) versus the full simultaneous PDF-SMEFT fit (orange). Both fits incorporate the complete global dataset.}
    \label{fig:simu_vs_fixed_bounds}
\end{figure}

In Fig.~\ref{fig:simu_vs_fixed_bounds}, we compare the final marginalized 95\% confidence bounds on the Wilson coefficients obtained 
from the full simultaneous fit against the fixed-PDF global baseline.

Previous studies have shown that allowing the PDFs and SMEFT coefficients to vary simultaneously can lead to a noticeable broadening of the Wilson-coefficient constraints. This effect is particularly relevant when high-mass Drell-Yan observables are included, since their tails can exhibit an intrinsic structural cross-talk: the neural-network PDF optimizer can adjust the high-$x$ valence and sea quark distributions to ``absorb'' mild high-energy deviations, effectively shielding the Wilson coefficients from being strictly bounded. 

Such effects have also been investigated in closure-test settings in which New Physics contributions are injected into pseudodata, demonstrating that the fitted PDFs can partially absorb the BSM-induced distortions. As a consequence, the SMEFT bounds obtained from a simultaneous PDF-SMEFT fit can differ from those derived using PDFs determined in a separate global PDF-only fit and held fixed in the SMEFT analysis.

In this analysis, Fig.~\ref{fig:simu_vs_fixed_bounds} shows that the bounds obtained in the simultaneous PDF–SMEFT fit are largely stable with respect to the fixed-PDF baseline. 
Most Wilson coefficients are essentially unchanged, but a handful show a moderate broadening: most notably $c_{lq}^{(3)}$, whose 95\% confidence interval widens by approximately 
a factor of two, in line with the findings of previous studies in the context of the HL-LHC~\cite{Iranipour:2022iak}. Some relaxation of the bounds is also observed for $c_{\varphi D}$ and $c_{\varphi ud}$. These are precisely the operators most sensitive to the high-energy Drell-Yan tails, where the neural-network PDF parametrisation retains some capacity 
to absorb new-physics-like distortions. Furthermore, the central values of several other coefficients shift by less than $1\sigma$ relative to the fixed-PDF baseline, reflecting a mild redistribution of constraining power in the simultaneous fit.

One possible explanation for the overall stability 
is that the inclusion of the low-energy precision observables, which are decoupled from high-$x$ PDF uncertainties, partially resolves 
potential degeneracies between the PDFs and the electroweak and four-fermion Wilson coefficients. However, the near-identical bounds 
for the majority of operators may also reflect an intrinsically weak PDF–SMEFT interplay for the particular operator basis and datasets considered here.

A particularly interesting exception to this overall stability is the right-handed charged-current coefficient $c_{\varphi ud}$. While both fits prefer a negative central value, the simultaneous PDF--SMEFT fit exhibits a deviation of more than $2\sigma$ from the SM expectation. A similar preference for right-handed charged-current interactions was reported in Ref.~\cite{Cirigliano:2023nol}, where the corresponding operator, denoted by $Q_{Hud}$, was shown to alleviate tensions in the SM description of semileptonic kaon and nuclear $\beta$-decay observables through contributions to vector and axial-vector transition amplitudes with opposite relative signs.

Although $C_{\phi ud}$ is directly constrained only by the low-energy datasets considered in this analysis, Fig.~\ref{fig:simu_vs_fixed_bounds} demonstrates that its marginalised bounds are 
nevertheless sensitive to the PDFs. In particular, performing the SMEFT fit using PDFs extracted from the full collider dataset under the SM assumption
shifts the bounds towards the SM expectation, such that the deviation of more than $2\sigma$ observed in the simultaneous SMEFT+PDF fit is no longer present. This highlights 
how potential BSM contamination of the PDFs can propagate through correlations between Wilson coefficients, even affecting operators to which the collider datasets have no 
direct sensitivity, and thereby potentially obscuring indications of new physics originating from low-energy measurements.

\begin{figure}[ht!]
    \centering
    \begin{subfigure}[b]{0.45\textwidth}
        \centering
        \includegraphics[width=\textwidth]{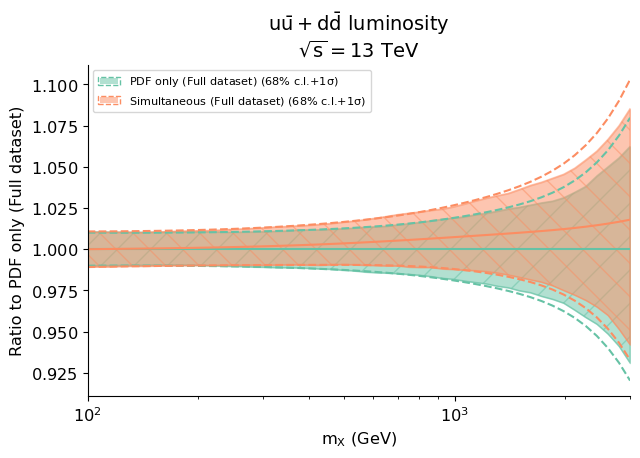} 
        \caption{}
        
    \end{subfigure}
    \hfill
    \begin{subfigure}[b]{0.45\textwidth}
        \centering
        \includegraphics[width=\textwidth]{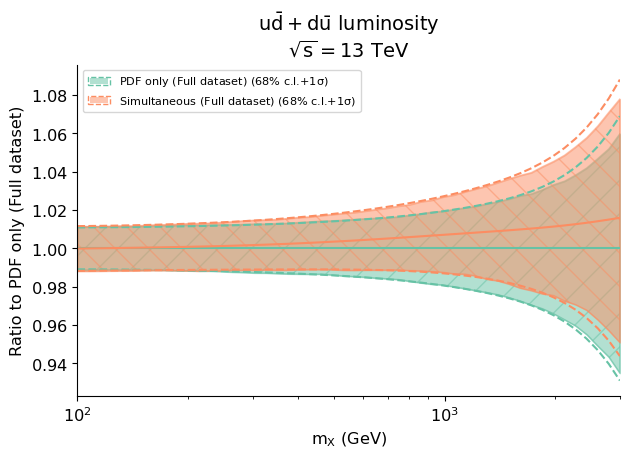} 
        \caption{}
        
    \end{subfigure}
    \caption{The neutral-current (a) and charged-current luminosity (b) evaluated at $\sqrt{s}=13$\,TeV as a function of the invariant mass $m_X$, comparing a pure Standard Model PDF fit (green) with the full simultaneous PDF--SMEFT fit (orange).}
    \label{fig:luminosities}
\end{figure}

To evaluate the impact of the SMEFT parameters on the extracted proton structure, we analyze the resulting parton-parton luminosities. 
Fig.~\ref{fig:luminosities} displays the  neutral-current (a) and charged-current luminosity (b) at $\sqrt{s} = 13$\,TeV as a 
function of the final-state invariant mass $m_X$, comparing a pure SM PDF fit ($c_i = 0$) against our full simultaneous PDF-SMEFT extraction. 

The comparison reveals a mild upward shift in the central values of both parton luminosities in the simultaneous PDF--SMEFT fit relative to the PDF-only baseline, which becomes more pronounced towards the multi-TeV high-mass tail ($m_X > 1\,\mathrm{TeV}$). These shifts nevertheless remain well within the corresponding PDF uncertainties across the entire kinematic range. While the uncertainties in both fits are approximately $1\%$ at low invariant masses and increase towards larger $m_X$, those obtained in the simultaneous fit exhibit only a mild broadening relative to the PDF-only baseline. This behaviour is consistent with the picture emerging from Fig.~\ref{fig:simu_vs_fixed_bounds}, whereby the simultaneous determination of the PDFs and Wilson coefficients introduces some additional freedom in the large-$x$ region without substantially degrading the precision of either determination. The modest shifts in the central values and the limited broadening of the luminosity uncertainties suggest that the PDF--SMEFT interplay remains relatively weak for the datasets and operator basis considered here. This may be partly attributed to the complementary constraints provided by the low-energy precision observables, which help to lift potential degeneracies between the PDFs and SMEFT coefficients.

\section{Conclusions}
\label{sec:conclusion}

In this work, we have presented a comprehensive global analysis of the Standard Model Effective Field Theory, 
establishing a unified framework that spans the low- and high-energy frontiers. By exploiting and extending the 
simultaneous fitting capabilities of the open-source \simunet{} architecture, we have performed a joint optimization of 
PDFs and a 15-dimensional space of dimension-six SMEFT Wilson coefficients. 

Our analysis introduces several key novelties to the global interpretation landscape. First, by systematically matching the 
SMEFT onto the Low-Energy Effective Field Theory (LEFT/WEFT), we integrated an extensive suite of low-energy precision inputs. 
This suite expands upon traditional nuclear $\beta$-decays and leptonic/semileptonic meson decays by incorporating, for the first time, neutral-current parity-violating observables (including PVES, APV, PVDIS, and CE$\nu$NS). Second, we completed the high-energy 
frontier by supplementing standard EWPOs, Higgs signal strengths, and diboson production data with the recently released ATLAS high-mass charged-current 
Drell-Yan dataset alongside neutral-current channels.

The numerical results of our fits demonstrate a powerful, multi-scale synergy. Within the fixed-PDF baseline, the low-energy precision data 
act as a strong parameter anchor, breaking persistent degeneracies among four-fermion and vertex correction operators that plague collider-only fits. 
Most notably, the inclusion of the ATLAS high-mass charged-current Drell-Yan data successfully collapsed a prominent flat direction 
in the $(c_{lq}^{(1)}, c_{lq}^{(3)})$ plane by isolating the triplet operator via $W^{\pm}$ exchange. Similarly, the addition of 
parity-violating datasets cleanly resolved the correlation between $c_{\phi q}^{(-)}$ and the right-handed quark vertex corrections ($c_{\phi u}$, $c_{\phi d}$).

When transitioning to the full simultaneous \simunet{} configuration, the inclusion of this multi-scale dataset provides a useful framework for 
assessing potential PDF-SMEFT cross-talk. While high-energy Drell-Yan tails in isolation can trade New Physics signatures for distortions in the 
large-$x$ quark luminosities, the low-energy precision observables offer a complementary and largely PDF-independent probe of the relevant SMEFT 
directions. These measurements may therefore play an important role in disentangling genuine SMEFT effects from high-$x$ proton structure. 
In this analysis, the resulting joint bounds remain stable compared to those obtained in the fixed-PDF reference, and the central values of 
the extracted parton-parton luminosities are stable across the full kinematic range, while the PDF uncertainties show a visible, though moderate, broadening 
at large invariant masses in the simultaneous fit compared to the fixed-PDF baseline. This behaviour is consistent with the low-energy 
data helping to reduce potential PDF-SMEFT degeneracies across the bulk of the kinematic range, although the residual broadening at high 
$M_X$ reflects the additional freedom retained by the neural-network PDF parametrisation in the large-$x$ region. 
It may also reflect an intrinsically weak PDF-SMEFT interplay for the particular operator basis and dataset considered here.

\medskip

The framework established in this paper opens several clear avenues for future investigation. First, we plan to investigate the 
dependence of the PV observables on PDFs, as done in Ref.~\cite{Whitehill:2026swa}, instead of using a fixed input PDF set. 
This is non-trivial as it would require the implementation of higher-twist corrections, as discussed in Ref.~\cite{Ball:2025xtj}. 
Moreover, we plan to expand our operator basis: to avoid unconstrained flat directions, this study limited its focus to 15 electroweak 
and four-fermion operators, and a natural next step is to extend the operator space to a fully general, flavour-anisotropic 
basis, tracking how simultaneous fits scale under a significantly larger parameter dimensionality. Along those lines, the integration 
of the top and jet sectors will be crucial: while the top-quark sector was omitted here to restrict the operator count, top and 
jet data are highly sensitive to four-quark and gluon operators, and a simultaneous fit will reveal whether their heavy PDF 
dependence causes similar cross-talk in the gluon and sea-quark channels. Furthermore, a forthcoming release of \simunet{} will 
make publicly available the full SMEFT likelihood for all observables entering the pure SMEFT fit, in a {\tt POPxf} 
format~\cite{Brivio:2025mww} compatible with the {\tt smelli} framework. This will enable users to combine the high-energy 
and low-energy constraints developed here with the broader {\tt smelli} likelihood, providing the community with comprehensive, 
validated likelihoods that span both the collider and precision frontiers. 
Finally, the methodology developed here is perfectly positioned to incorporate data from upcoming high-luminosity facilities. 
Incorporating the final HL-LHC data streams, precision parity-violating measurements from the Electron-Ion Collider 
(EIC)~\cite{Boughezal:2021kla}, and ultra-precise coherent neutrino scattering measurements will further sharpen the 
joint determination of hadron structure and BSM physics.

In conclusion, this work demonstrates that the path to a robust, model-independent search for New Physics does not rely on high-energy colliders or 
low-energy precision experiments in isolation, but rather on their self-consistent, simultaneous synthesis with the internal structure of the nucleon.

\section*{Acknowledgments}
E.~C. is
supported by Newnham College, Cambridge and by the STFC Centre for Doctoral Training
(CDT) in Data Intensive Science. 
M.~U. is supported by the European Research Council under the European Union’s Horizon 2020 research and innovation Programme (grant agreement n.950246), 
and partially by the STFC consolidated grant ST/T000694/1 and ST/X000664/1. 
T.~T. is supported by his four cats.

\newpage
\appendix

\section{Detailed information on the data included in the analysis}
\label{app:datasets}

In this appendix, we summarise the details of the datasets included in our global SMEFT and PDF determination.

\subsection{Collider data}

We describe each category in turn.

\noindent {\bf EWPOs:}
In Table~\ref{tab:dataset_EWPO}, we list EWPOs included in this study. 
The dataset includes the pseudo-observables
measured on the $Z$ resonance by LEP~\cite{ALEPH:2005ab}, 
which include cross sections, forward-backward and polarised asymmetries. 
Branching ratios of the decay of the $W$ boson into leptons~\cite{ALEPH:2013dgf} are also included, 
along with the LEP measurement of the Bhabha scattering cross section~\cite{ALEPH:2013dgf}. 
Finally we include the measurement of the effective electromagnetic coupling constant~\cite{ParticleDataGroup:2022pth}, 
for a total of 44 data points.  These datasets and their SMEFT predictions
are taken from the public \smefit{} code~\cite{Giani:2023gfq}.  These data points are all PDF independent, hence they directly 
affect only the SMEFT part of the fits.
\begin{table}[htb!]
    \begin{center}
  {\fontsize{8pt}{8pt}\selectfont
    \centering
     \renewcommand{\arraystretch}{2}
     \setlength{\tabcolsep}{5pt}
     \begin{tabular}{lcccccc}
       \toprule \textbf{Exp.}   & $\bf{\sqrt{s}}$ \textbf{(TeV)}
      &  \textbf{Observable} & $\mathcal{L}$ (fb${}^{-1}$) & $\mathbf{n_{\rm dat}}$ & \textbf{Dataset name}
       &\textbf{Ref.}\\
      \toprule
        \bf{LEP}
        & 0.250
        & Z observables
        & 
        & 19
        & LEP\_ZDATA
        & \cite{ALEPH:2005ab}\\
      \midrule
        \bf{LEP}
        & 0.196
        & $\mathcal{B}(W \rightarrow l^{-} \bar{v}_l)$
        & 3
        & 3
        & LEP\_BRW
        & \cite{ALEPH:2013dgf}\\
      \midrule
        \bf{LEP}
        & 0.189
        & $\sigma(e^+ e^- \rightarrow e^+ e^-)$
        & 3
        & 21
        & LEP\_BHABHA
        & \cite{ALEPH:2013dgf}\\
      \midrule
        \bf{LEP}
        & 0.209
        & $\hat{\alpha}^{(5)}_{\rm}(M_Z)$
        & 3
        & 1
        & ALPHAEW\_WITHTHUNC 
        & \cite{ParticleDataGroup:2022pth}\\
  
  \bottomrule
     \end{tabular}
     \vspace{0.3cm}
     \caption{Measurements of electroweak precision observables included in \simunet{}. The columns contain information on the 
     centre-of-mass energy, the observable, the integrated luminosity, the number 
     of data points, the dataset name as implemented in \simunet \ and the source.
      \label{tab:dataset_EWPO}
  }
  }
  \end{center}
  \end{table}

\vspace*{0.4cm}
\noindent {\bf Higgs sector:}
In Table~\ref{tab:dataset_Higgs}, we list the Higgs sector datasets included in this study. 
The Higgs dataset at the LHC includes the combination of Higgs signal strengths
by ATLAS and CMS for Run 1, and for Run 2 both signal strengths and STXS
measurements are used. ATLAS in particular provides the combination of stage 1.0 STXS bins
for the $4l$, $\gamma\gamma$, $WW$, $\tau\tau$ and $b\bar{b}$ decay channels, while for CMS we use the combination
of signal strengths of the $4l$,
$\gamma\gamma$, $WW$, $\tau^-\tau^+$, $b\bar{b}$ and $\mu^-\mu^+$ decay channels. 
We also include the $H \rightarrow Z\gamma$ 
and $H \rightarrow \mu^-\mu^+$ signal strengths from ATLAS, for a total of 73 data points. 
The Run I and CMS datasets and their corresponding predictions are taken from the \smefit{} code \cite{Giani:2023gfq},
whereas the STXS observables and signal strength measurements of
$H \rightarrow Z\gamma$ and $H \rightarrow \mu^-\mu^+$ are taken from the \fitmaker ~\cite{Ellis:2020unq} code.
The signal strength's dependence on the PDFs is almost completely negligible, given that the 
PDF dependence cancels in the ratio, hence all the data points are treated as PDF independent and 
are directly affected only by the SMEFT Wilson coefficients.
\begin{table}[htb!]
    \begin{center}
  {\fontsize{8pt}{8pt}\selectfont
    \centering
     \renewcommand{\arraystretch}{2}
     \setlength{\tabcolsep}{5pt}
     \begin{tabular}{lcccccc}
       \toprule \textbf{Exp.}   & $\bf{\sqrt{s}}$ \textbf{(TeV)}
      &  \textbf{Observable} & $\mathcal{L}$ (fb${}^{-1}$) & $\mathbf{n_{\rm dat}}$ & \textbf{Dataset name}
       &\textbf{Ref.}\\
      \toprule
        \bf{ATLAS and CMS}
        & 7 and 8
        & $\mu_{H \rightarrow \mu^+ \mu^-}$
        & 5 and 20
        & 22
        & ATLAS\_CMS\_SSinc\_RunI
        & \cite{ATLAS:2016neq}\\
      \midrule
        \bf{CMS}
        & 13
        & $\mu_{H}$
        & 35.9
        & 24
        & CMS\_SSINC\_RUNII
        & \cite{CMS:2018uag}\\
      \midrule
        \bf{ATLAS}
        & 13
        & $\mu_{H}$
        & 80
        & 25
        & ATLAS\_STXS\_RUNII
        & \cite{ATLAS:2019nkf}\\
      \midrule
        \bf{ATLAS}
        & 13
        & $\mu_{H \rightarrow Z \gamma}$
        & 139
        & 1
        & ATLAS\_SSINC\_RUNII\_ZGAM
        & \cite{ATLAS:2020qcv}\\
      \midrule
        \bf{ATLAS}
        & 13
        & $\mu_{H \rightarrow \mu^+ \mu^-}$
        & 139
        & 1
        & ATLAS\_SSINC\_RUNII\_MUMU
        & \cite{ATLAS:2020fzp}\\

  \bottomrule
     \end{tabular}
     \vspace{0.3cm}
     \caption{Same as~\ref{tab:dataset_EWPO} for the measurements of the Higgs sector.
      \label{tab:dataset_Higgs}  }
  }
  \end{center}
  \end{table}

\vspace*{0.4cm}
\noindent {\bf Diboson sector:}
In Table~\ref{tab:dataset_diboson}, we list the diboson sector datasets included in this study. 
We have implemented four measurements from LEP at low energy, three from ATLAS and one from CMS both at 13 TeV. 
The LEP measurements are of the differential 
cross-section of $e^+ e^- \rightarrow W^+ W^-$ as a function of the cosine of the $W$-boson polar angle. 
The ATLAS measurements are of 
the differential cross-section of $W^+ W^-$ as a function of the invariant mass of the electron-muon pair 
and of the transverse mass of the $W$-boson. 
The third ATLAS measurement is of the differential cross-section of $Zjj$ as a function of the azimuthal angle 
between the two jets, $\Delta \phi_{jj}$.  Although this is not a diboson 
observable, it is grouped here as this observable constrains operators typically 
constrained by the diboson sector,
such as the triple gauge interaction $O_{WWW}$. The CMS dataset is a measurement of the differential cross-section of $WZ$ as 
a function of the transverse momentum 
of the $Z$-boson, for a total of 82 data points. 
Almost all datasets and corresponding predictions in this sector are taken from the \smefit{} code~\cite{Giani:2023gfq},
except for the ATLAS measurement of $Zjj$ production, taken from the \fitmaker ~\cite{Ellis:2020unq} code.
While the LEP measurements are PDF-independent by definition, 
the LHC measurements of diboson cross sections do depend on PDFs. However, their impact on the 
PDFs is extremely mild, due to much more competitive constraints in the same Bjorken-$x$ region given by other measurements
such as single boson production and single boson production in association with jets. 
Therefore we consider all diboson measurements 
to be PDF-independent, i.e. the PDFs are fixed and their parameters are not varied in the computation of these observables. 
\begin{table}[htb!]
    \begin{center}
  {\fontsize{8pt}{8pt}\selectfont
    \centering
     \renewcommand{\arraystretch}{2}
     \setlength{\tabcolsep}{5pt}
     \begin{tabular}{lcccccc}
       \toprule \textbf{Exp.}   & $\bf{\sqrt{s}}$ \textbf{(TeV)}
      &  \textbf{Observable} & $\mathcal{L}$ (fb${}^{-1}$) & $\mathbf{n_{\rm dat}}$ & \textbf{Dataset name}
       &\textbf{Ref.}\\
      \toprule
        \bf{LEP}
        & 0.182
        & $d \sigma _{WW} / d cos(\theta _W)$
        & 0.164
        & 10
        & LEP\_EEWW\_182GEV
        & \cite{ALEPH:2013dgf}\\
      \midrule
        \bf{LEP}
        & 0.189
        & $d \sigma _{WW} / d cos(\theta _W)$
        & 0.588
        & 10
        & LEP\_EEWW\_189GEV
        & \cite{ALEPH:2013dgf}\\
      \midrule
        \bf{LEP}
        & 0.198
        & $d \sigma _{WW} / d cos(\theta _W)$
        & 0.605
        & 10
        & LEP\_EEWW\_198GEV
        & \cite{ALEPH:2013dgf}\\
      \midrule
        \bf{LEP}
        & 0.206
        & $d \sigma _{WW} / d cos(\theta _W)$
        & 0.631
        & 10
        & LEP\_EEWW\_206GEV
        & \cite{ALEPH:2013dgf}\\
      \midrule
        \bf{ATLAS}
        & 13
        & $d \sigma _{W^+W^-}/d m_{e \mu}$
        & 36.1
        & 13
        & ATLAS\_WW\_13TeV\_2016\_MEMU
        & \cite{ATLAS:2019rob}\\
      \midrule
        \bf{ATLAS}
        & 13
        & $d \sigma _{WZ} / d m_{T}$
        & 36.1
        & 6
        & ATLAS\_WZ\_13TeV\_2016\_MTWZ
        & \cite{ATLAS:2019bsc}\\
      \midrule
        \bf{ATLAS}
        & 13
        & $d \sigma(Zjj)/d \Delta \phi_{jj}$
        & 139
        & 12
        & ATLAS\_Zjj\_13TeV\_2016
        & \cite{ATLAS:2020nzk}\\
      \midrule
        \bf{CMS}
        & 13
        & $d \sigma _{WZ} / d p_{T}$
        & 35.9
        & 11
        & CMS\_WZ\_13TeV\_2016\_PTZ
        & \cite{ATLAS:2020nzk}\\

  \bottomrule
     \end{tabular}
     \vspace{0.3cm}
     \caption{Same as~\ref{tab:dataset_EWPO} for the measurements of the diboson sector.
      \label{tab:dataset_diboson}
  }
  }
  \end{center}
  \end{table}

\vspace*{0.4cm}
\noindent {\bf Drell-Yan sector:}
In Table~\ref{tab:dataset_drellyan}, we list the Drell-Yan sector datasets included in this study.
This sector includes fixed-target data from the E605, E886 and E906 experiments, Tevatron measurements from CDF and D0, and LHC measurements from ATLAS, CMS, and LHCb.
These datasets are taken from the \nnpdf{} dataset. Table~\ref{tab:dataset_HMDY} lists the datasets included from the high-mass Drell Yan sector. This extends \nnpdf{} with the inclusion of two single-differential CMS datasets~\cite{CMS:2014jea, CMS:2018mdl} and the charged-current ATLAS dataset~\cite{ATLAS:2025hhn}, adding a total of 96 data points. In total, the high-mass Drell Yan sector provides 259 data points, all contributing to both PDF and SMEFT constraints.

\begin{table}[t!]
    \vspace*{-0.5cm}
    \begin{center}
  {\fontsize{8pt}{8pt}\selectfont
     \renewcommand{\arraystretch}{2}
     \setlength{\tabcolsep}{5pt}
     \begin{tabular}{lcccccc}
       \toprule \textbf{Exp.}   & $\bf{\sqrt{s}}$ \textbf{(TeV)}
      &  \textbf{Observable} & $\mathcal{L}$ (fb${}^{-1}$) & $\mathbf{n_{\rm dat}}$ & \textbf{Dataset name}
       &\textbf{Ref.}\\
      \toprule
        \bf{E886}
        & 0.8
        & $d\sigma^d_{DY}/d\sigma^p_{DY}$
        & ?
        & 15
        & DYE886R\_dw\_ite
        & \cite{NuSea:2001idv} \\
        \midrule
        \bf{E886}
        & 0.8
        & $d\sigma^p_{DY}/dy dm_{ll}$
        & ?
        & 89
        & DYE886P 
        & \cite{NuSea:2003qoe, Webb:2003bj} \\
        \midrule
        \bf{E605}
        & 0.04
        & $d\sigma^p_{DY}/dx_F dm_{ll}$
        & ?
        & 85
        & DYE605\_dw\_ite
        & \cite{Moreno:1990sf} \\
        \midrule
        \bf{E906}
        & 0.015
        & $\sigma^d_{\rm DY}/\sigma^p_{\rm DY}$
        & ?
        & 6
        & DYE906R\_dw\_ite
        &  \cite{SeaQuest:2021zxb}\\
        \midrule
        \bf{CDF}
        & 1.96
        & $d\sigma_Z/dy_Z$
        & 2.1
        & 28
        & CDFZRAP\_NEW
        & \cite{CDF:2010vek} \\
        \midrule
        \bf{D0}
        & 1.96
        & $d\sigma_Z/dy_Z$
        & 0.4
        & 28
        & D0ZRAP\_40
        & \cite{D0:2007djv} \\
        \midrule
        \bf{D0}
        & 1.96
        &  $d\sigma_{W\rightarrow\mu\nu}/d\eta_\mu$ asy
        & 7.3
        & 9 
        & D0WMASY
        & \cite{D0:2013xqc} \\
        \midrule
        \bf{ATLAS}
        & 7
        & $d\sigma_W/d\eta_l,d\sigma_Z/dy_z$
        & 0.035
        & 30
        & ATLASWZRAP36PB
        & \cite{ATLAS:2011qdp}\\
        \midrule
        \bf{ATLAS}
        & 7
        &  $d\sigma_{Z\rightarrow e^+e^-}/dm_{e^+e^-}$
        & 0.035
        & 6
        & ATLASLOMASSDY11EXT
        & \cite{ATLAS:2014ape}\\
        \midrule
        \bf{ATLAS}
        & 7
        &  $d\sigma_W/d\eta_l,d\sigma_Z/dy_z$
        & 4.6
        & 61
        & ATLASWZRAP11
        & \cite{ATLAS:2016nqi}\\
        \midrule
        \bf{ATLAS}
        & 8
        &  $d^2\sigma/dm_{\ell\ell}d|y_{\ell\ell}|$
        & 20.1
        & 60
        & ATLAS\_DY\_2D\_8TEV\_LOWMASS
        & \cite{ATLAS:2017rue}\\
        \midrule
        \bf{ATLAS}
        & 13
        & $\sigma^{fid}$
        & 0.081
        & 3
        & ATLAS\_WZ\_TOT\_13TEV
        & \cite{ATLAS:2016fij}\\
        \midrule
        \bf{ATLAS}
        & 8
        & $d\sigma/dp_T^{W^+}$, $d\sigma/dp_T^{W^-}$
        & 20.2
        & 15, 15
        & ATLAS\_(WP, WM)\_JET\_8TEV\_PT
        & \cite{ATLAS:2017irc}\\
        \midrule
        \bf{ATLAS}
        & 8
        & $d\sigma_{Z/\gamma^{*}}/dp_T$
        & 20.3
        & 44, 48
        & ATLASZPT8TEV(M,Y)DIST
        & \cite{ATLAS:2015iiu}\\
        \midrule
        \bf{ATLAS}
        & 8
        & $d^2\sigma/dp_Td|y|$
        & 20.2
        & 171
        & ATLAS\_1JET\_8TEV\_R06\_DEC
        & \cite{ATLAS:2017kux}\\
        \midrule
        \bf{ATLAS}
        & 7
        & $d^2\sigma/dm_{12}dy*$
        & 3.5
        & 90
        & ATLAS\_2JET\_7TEV\_R06
        & \cite{ATLAS:2013jmu}\\
        \midrule
        \bf{ATLAS}
        & 13
        & $d\sigma_{\gamma}/dp_T ^\gamma$
        & 3.2
        & 53
        & ATLASPHT15\_SF
        & \cite{ATLAS:2017nah}\\
        \midrule
        \bf{CMS}
        & 7
        &  $dA_{e} / d|y|$
        & 0.84
        & 11
        & CMSWEASY840PB 
        & \cite{Chatrchyan_2012} \\
        \midrule
        \bf{CMS}
        & 7
        & $dA_{\mu} / d|y|$
        & 4.7
        & 11
        & CMSWMASY47FB
        & \cite{CMS:2013pzl} \\
        \midrule
        \bf{CMS}
        & 8
        &  $d\sigma_{W\rightarrow \mu\nu}/d\eta_\mu$
        & 18.8
        & 22
        & CMSWMU8TEV
        & \cite{CMS:2013pzl} \\
        \midrule
        \bf{CMS}
        & 8
        &  $d\sigma_Z/dp_T$
        & 19.7
        & 28
        & CMSZDIFF12
        & \cite{Khachatryan_2015} \\
        \midrule
        \bf{CMS}
        & 7
        & $d^2\sigma/dm_{12}dy$
        & 5.0
        & 54
        & CMS\_2JET\_7TEV
        & \cite{CMS:2012ftr}\\
        \midrule
        \bf{CMS}
        & 8
        & $d^2\sigma/dp_Tdy$
        & 19.7
        & 185
        & CMS\_1JET\_8TEV
        & \cite{CMS:2016lna}\\
        \midrule
        \bf{LHCb}
        & 7
        & $d\sigma / d|y|$
        & 0.94
        & 9
        & LHCBZ940PB
        & \cite{LHCb:2012gii}\\
        \midrule
        \bf{LHCb}
        & 8
        & $d\sigma / d|y|$
        & 2.0
        & 17
        & LHCBZEE2FB\_40
        & \cite{LHCb:2015kwa}\\
        \midrule
        \bf{LHCb}
        & 7
        & $d\sigma / d|y|$
        & 1.0
        & 29
        & LHCBWZMU7TEV
        & \cite{LHCb:2015okr}\\
        \midrule
        \bf{LHCb}
        & 8
        & $d\sigma / d|y|$
        & 2.0
        & 30
        & LHCBWZMU8TEV
        & \cite{LHCb:2015mad}\\
        \midrule
        \bf{LHCb}
        & 13
        & $d\sigma / d|y|$
        & 0.294
        & 16, 15
        & LHCB\_Z\_13TEV\_DI(MUON, ELECTRON)
        & \cite{LHCb:2016fbk}\\
  \bottomrule
     \end{tabular}
     \vspace{0.3cm}
     \caption{Same as~\ref{tab:dataset_EWPO} for the measurements of the Drell Yan sector.
      \label{tab:dataset_drellyan}  }
  }
  \end{center}
  \end{table}

\begin{table}[htb!]
    \begin{center}
  {\fontsize{8pt}{8pt}\selectfont
    \centering
     \renewcommand{\arraystretch}{2}
     \setlength{\tabcolsep}{5pt}
     \begin{tabular}{lcccccc}
       \toprule \textbf{Exp.}   & $\bf{\sqrt{s}}$ \textbf{(TeV)}
      &  \textbf{Observable} & $\mathcal{L}$ (fb${}^{-1}$) & $\mathbf{n_{\rm dat}}$ & \textbf{Dataset name}
       &\textbf{Ref.}\\
      \toprule
        \bf{ATLAS}
        & 8
        & $d^2\sigma/dm_{\ell\ell}d|y_{\ell\ell}|$
        & 20.3
        & 48
        & ATLASDY2D8TEV
        & \cite{ATLAS:2016gic}\\
      \midrule
        \bf{ATLAS}
        & 7
        & $d\sigma_{Z/\gamma^{*}}/dM_{ll}$ 
        & 4.9
        & 5
        & ATLASZHIGHMASS49FB
        & \cite{ATLAS:2013xny}\\
      \midrule
        \bf{ATLAS}
        & 13
        & ${d \sigma}/{d m_T^W}$ 
        & 140
        & 12
        & ATLAS\_WPWM\_13TEV\_HMT\_DIF-LEP-PM
        & \cite{ATLAS:2025hhn}\\
      \midrule
        \bf{CMS}
        & 7
        & $d\sigma_{Z/\gamma^{*}}/dy$
        & 4.5
        & 110
        & CMSDY2D11
        & \cite{CMS:2013zfg}\\
      \midrule
        \bf{CMS}
        & 8
        & $d\sigma/dm_{ll}$
        & 19.7
        & 41
        & CMSDY1D12
        & \cite{CMS:2014jea}\\
      \midrule
        \bf{CMS}
        & 13
        & $d\sigma/dm_{ll}$
        & 2.8
        & 43
        & CMS\_HMDY\_13TEV
        & \cite{CMS:2018mdl}\\
  \bottomrule
     \end{tabular}
     \vspace{0.3cm}
     \caption{Same as~\ref{tab:dataset_EWPO} for the measurments of the high-mass Drell Yan sector.
      \label{tab:dataset_HMDY}
  }
  }
  \end{center}
  \end{table}

\subsection{Low-energy data}
\label{app:low-energy-data}

\noindent\textbf{Effective uncertainties for charged-current observables.}
The global analysis in Ref.~\cite{Cirigliano:2023nol} fits the CKM parameter $\lambda$, hadronic inputs and nuclear nuisance parameters, including $\eta_{1,2,3}$ for radiative and nuclear-structure corrections to $\beta$ decays. We use the pulls from its Standard Model fit to set the weights of the charged-current data in \simunet{}, where the low-energy parameters are held fixed.

For each retained charged-current observable $i$, let $D_i$ be its measured value, $T_i^{\rm SM}$ its fixed \simunet{} SM prediction, and $p_i^{\rm SM}$ its pull at the SM best fit of Ref.~\cite{Cirigliano:2023nol}. We assign the effective uncertainty
\begin{equation}
    \sigma_{i,\mathrm{eff}} = \left\lvert \frac{D_i-T_i^{\rm SM}}{p_i^{\rm SM}} \right\rvert \,.
\end{equation}
This uncertainty replaces the experimental error in the fit and is held fixed as the Wilson coefficients vary. The construction reproduces the magnitude of each retained SM pull.

We omit $\mathcal{F}t({}^{17}\mathrm{F})$ from the fit. Its near-zero SM pull yields an extremely large effective uncertainty, giving the measurement negligible weight under this prescription. The parity-violating data described below are included following Ref.~\cite{Crivellin:2021bkd}.

\vspace*{0.4cm}
\noindent {\bf Neutron and nuclear $\beta$-decays:} 
We include measurements of $0^+\to 0^+$ transitions, neutron decay, as well as decay correlation measurements, following Ref.~\cite{Cirigliano:2023nol}. For measurements of neutron decays, the observables include the lifetime, as well as several correlation coefficients, $\tilde A_n$, $\tilde B_n$, $ a_n$, $\tilde a_n$, and $\lambda_{AB}$. 
The included ${\cal F} t$ values from superallowed decays are listed in Table~\ref{tab:superallowed}, while the correlation measurements used in the nuclei are collected in Table~\ref{tab:NuclearData}. For definitions of the observables, see Ref.~\cite{Falkowski:2020pma}. Finally, we include the measurement of the ratio of the axial and vector form factors in $\Lambda\to p e\nu_e$~\cite{ParticleDataGroup:2022pth}.

The theoretical expressions of these observables do not just involve the SM and SMEFT parameters, but also depend on several hadronic and nuclear matrix elements. The relevant hadronic input for the SM consists of the vector and axial nucleon charges, $g_V$ and $g_A$, and radiative corrections, $\Delta_R^{V,A}$, while the contributions due to BSM interactions require the scalar and tensor charges, $g_S$, $g_T$. Pseudoscalar interactions are suppressed by $\mathcal{O}(m_e \Lambda_\chi/m_\pi^2)$, and we neglect their contributions to $\beta$-decays.
We summarize these hadronic parameters in the first column of Table~\ref{tab:hadronic}. 

\begin{table}[ht]
\footnotesize
\begin{center}
\begin{tabular}{ ccccccccccccccccc}
\hline\hline
$^{10}$C   & 
$^{14}$O   & 
$^{22}$Mg  &
$^{26m}$Al &
$^{26}$Si & 
$^{34}$Cl  &
$^{34}$Ar  & 
$^{38m}$K  & 
$^{38}$Ca  & 
$^{42}$Sc  & 
$^{46}$V   & 
$^{50}$Mn  & 
$^{54}$Co  & 
$^{62}$Ga  & 
$^{74}$Rb  \\
\hline\hline
\end{tabular}
\end{center}
\caption{
The parent nuclei participating in superallowed decays for which we use their measured ${\cal F}t$ values in the fits, following Refs.~\cite{Falkowski:2020pma,Hardy2020Oct}.}
\label{tab:superallowed}
\end{table}

\begin{table}[ht]
\begin{center}
\begin{tabular}{
l@{\hspace{5mm}} c @{\hspace{3mm}}  c@{\hspace{3mm}}  c@{\hspace{3mm}}  c@{\hspace{3mm}}
 r@{\hspace{5mm}}  
c@{\hspace{3mm}}  c 
}
\hline\hline
Parent		& $J_i$	& $J_f$	& Type	& Observable 
& Ref.
\\ \hline
$^6$He		& 0		& 1		& GT/$\beta^-$		& $\tilde{a}$	& \cite{Johnson:1963zza}
\\
$^{32}$Ar		& 0		& 0		&  F/$\beta^+$		& $\tilde{a}$	& \cite{Adelberger:1999ud} 
\\
$^{38m}$K	& 0		& 0		&  F/$\beta^+$		& $\tilde{a}$	& \cite{Gorelov:2004hv}
\\
$^{60}$Co		& 5		& 4		& GT/$\beta^-$		& $\tilde{A}$		& \cite{Wauters:2010gh} \\
$^{67}$Cu		& 3/2		& 5/2		& GT/$\beta^-$		& $\tilde{A}$	& \cite{Soti:2014xua} 
\\
$^{114}$In	& 1		& 0		& GT/$\beta^-$		& $\tilde{A}$	& \cite{Wauters:2009jw}
\\
$^{14}$O/$^{10}$C &	&		& F-GT/$\beta^+$	& $P_F/P_{GT}$ 	&  \cite{Carnoy:1991jd}
\\
$^{26}$Al/$^{30}$P &	&		& F-GT/$\beta^+$	& $P_F/P_{GT}$    & \cite{Wichers:1986es}
\\ \hline\hline
\end{tabular}
\end{center}
\caption{
Data from correlation measurements in pure Fermi and pure Gamow-Teller decays used in the fits. Table adapted from~\cite{Falkowski:2020pma}.}
\label{tab:NuclearData}
\end{table}
\begin{table}[ht]
\footnotesize
\begin{center}
\begin{tabular}{ |llc|llc|}
\hline\hline
\multicolumn{3}{|c|}{Baryon decays} & \multicolumn{3}{c|}{$K$ and $\pi$ decays}\\\hline
$g_A $ &  $1.246(28)$ & \cite{FlavourLatticeAveragingGroupFLAG:2021npn} & $f_+(0)$ & $0.9698(17)$&\cite{FlavourLatticeAveragingGroupFLAG:2021npn} \\
$g_S$& $1.02(10)$ & \cite{FlavourLatticeAveragingGroupFLAG:2021npn} & $\frac{f_{K^\pm}}{f_{\pi^\pm}}$ & 1.1932(21)& \cite{FlavourLatticeAveragingGroupFLAG:2021npn}\\ 
$g_T $& $0.989(34)$ & \cite{FlavourLatticeAveragingGroupFLAG:2021npn}& $f_{K^\pm}$ &$155.7(3)\, {\rm MeV}$ & \cite{FlavourLatticeAveragingGroupFLAG:2021npn}\\
$\Delta_R^V$ & $0.02467(22)$& \cite{Seng:2018yzq}& RC$_\pi$ &$0.0332(3)$& \cite{Feng:2020zdc}\\
$\frac{\Delta_R^A-\Delta_R^V}{2}$ & $0.00013(13)$& \cite{Gorchtein:2021fce}& $c_1^\pi$ & $-2.4(5)$ & \cite{DescotesGenon:2005pw} \\
$g_1/f_1\Big|_{\Lambda\to p}$& $0.72(7)$& \cite{Gonzalez-Alonso:2016etj,QCDSFUKQCD:2010cha}&$\Delta_{\rm CT}$& $-0.0035(80)$ & \cite{FlaviaNetWorkingGrouponKaonDecays:2010lot}\\
&&&$\frac{B_T(0)}{f_+(0)}$&$0.68(3)$ & \cite{Baum:2011rm}\\
\hline\hline
\end{tabular}
\end{center}
\caption{The hadronic input that enters $\beta$-decays. We approximate $g_V=1$, which holds with high precision. Whenever appropriate, the value on a renormalization scale $\mu=2$ GeV is given.}
\label{tab:hadronic}
\end{table}

\vspace*{0.4cm}
\noindent {\bf Kaon and pion decays:}
The experimental data used for the decays of pion and kaon are collected in Table~\ref{tab:mesons}. Apart from the two-body decay rates (the first column), we include determinations of $V_{us}^{\ell}f_+(0)$ from the $K\to \pi e\nu$ and $K\to \pi \mu\nu$ rates and measurements of the differential distributions (second column). We follow~\cite{Gonzalez-Alonso:2016etj} for the theoretical expressions of the observables and summarize the relevant hadronic input in the second column of Table~\ref{tab:hadronic}.

\begin{table}[ht]
\footnotesize
\begin{center}
\begin{tabular}{ |lll|llc|}
\hline\hline 
\multicolumn{3}{|c}{$P_{l2}$} & \multicolumn{3}{|c|}{$K_{l3}$} \\ \hline
Observable & value & Ref. & Observable & value & Ref. \\ \hline
$ R_\pi$ &$ 1.2344(30)\cdot 10^{-4}$ &\cite{PiENu:2015seu} &$|\tilde V_{us}^{e}f_+^K(0)| $ & $ 0.21626(40)$ & \cite{Seng:2021nar}\\
$ R_K $ &$ 2.488(9)\cdot 10^{-5}$ & \cite{ParticleDataGroup:2022pth}&$|\tilde V_{us}^{\mu}f_+^K(0)| $ & $ 0.21667(52)$ & \cite{Seng:2021nar}\\
$ \Gamma(K_{\mu2}) $ &$ 5.134(10)\cdot 10^{-7}s^{-1}$ &\cite{ParticleDataGroup:2022pth}& $\log C $& $0.1985(70)$ & \cite{Moulson:2017ive}\\ 
$ {\rm Br}(\pi_{\mu2}) $ &$ 0.9998770(4)$ &\cite{ParticleDataGroup:2022pth}&$2\epsilon_T^{s\mu}\frac{B_T(0)}{f_+(0)}$& $0.0007(71)$ & \cite{Yushchenko:2003xz}\\
$ \tau_{\pi^+} $ &$ 2.6033(5)\cdot 10^{-8}s$ &\cite{ParticleDataGroup:2022pth}&&&\\
\hline\hline
\end{tabular}
\end{center}
\caption{
The experimental input used for kaon and $\pi$ decays. Note that the determinations of $|\tilde V_{us}^{\ell}f_+^K(0)|$ involve a correlation that is not shown in the Table. The ratio $R_\mu=\Gamma(K_{\mu 2})/\Gamma(\pi_{\mu 2})$, with $\Gamma(\pi_{\mu 2})={\rm Br}(\pi_{\mu 2})/\tau_{\pi^+}$, is derived from the Table.}
\label{tab:mesons}
\end{table}

\vspace*{0.4cm}
\noindent {\bf Parity Violation Observables:}
We briefly list the parity-violating observables included in our analysis, which closely follow the study in Ref.~\cite{Crivellin:2021bkd}.
\begin{itemize}
    \item Parity-Violating Electron Scattering (PVES): we include the weak charge of the proton, $Q_\text{w}^p$, measured by the $Q_\text{weak}$ experiment at Jefferson Lab~\cite{Qweak:2018tjf,Carlini:2019ksi}.
    
    \item Atomic Parity Violation (APV): we include the extraction of the cesium weak charge, $Q_\text{w}(^{133}\text{Cs})$, in Ref.~\cite{Cadeddu:2021dqx}.

    \item Parity-Violating Deep Inelastic Scattering (PVDIS): we include two measurements from the Jefferson Lab PVDIS collaboration~\cite{PVDIS:2014cmd,Wang:2014guo}, for the two kinematic settings with a deuterium target.

    \item Coherent Elastic Neutrino-Nucleus Scattering (CE$\nu$NS): we include two measurements by COHERENT for CsI (with a CsI[Na] detector)~\cite{COHERENT:2018gft} and Ar (with a liquid argon detector)~\cite{COHERENT:2020iec}.
    
\end{itemize}

\newpage
\clearpage
\bibliographystyle{JHEP}
\bibliography{references.bib}

\end{document}